\documentclass[12pt]{iopart} \usepackage{hyperref} \usepackage{iopams}
\usepackage{epsfig}
\def\lsi{\raise0.3ex\hbox{$<$\kern-0.75em\raise-1.1ex\hbox{$\sim$}}}
\def\gsi{\raise0.3ex\hbox{$>$\kern-0.75em\raise-1.1ex\hbox{$\sim$}}}
\newcommand{\ls}{\mathop{\lsi}} \newcommand{\gs}{\mathop{\gsi}}

\begin{document}

\begin{flushright}
DAMTP-2005-6\\
31 January 2005
\end{flushright}

\title{The end of locked inflation} \author{E. J. Copeland\dag\
and
A. Rajantie\ddag} 

\address{\dag\ School of Physics and Astronomy,
University of Nottingham, University Park, Nottingham, NG7 2RD,
United Kingdom}

\address{\ddag\ DAMTP, University of Cambridge, Wilberforce Road,
Cambridge CB3 0WA, United Kingdom} 

\ead{a.k.rajantie@damtp.cam.ac.uk
\\~~~~~~~ed.copeland@nottingham.ac.uk}

\begin{abstract}
We investigate the end of the inflationary period in the 
recently proposed scenario of
locked inflation, and consider various constraints
arising from density perturbations, loop
corrections, parametric resonance and defect formation. We show that
in a scenario where there is one long period of locked
inflation,   it is not possible to satisfy all of these constraints
without having a period of saddle inflation afterwards, which would
wipe out all observable signatures. On the other hand, if one does not
insist on satisfying the loop correction constraint, saddle inflation
can be avoided, but then inflation must have ended through parametric
resonance.
\end{abstract}
\section{Introduction}

For a paradigm that has become so universally accepted as the one
responsible for the generation of the primordial density fluctuations
in the universe, as well as the observed homogeneity,  isotropy  and
spatial flatness, it is remarkable that the Inflationary Universe
Scenario appears to sit so uncomfortably in most current
particle physics models. Following the initial work of Guth
\cite{Guth:1980zm}, where he demonstrated how old inflation could
occur, it was soon realised  that inflation could not
end gracefully. The next variant,  new inflation, where the false
vacuum region of the potential was replaced by a slowly rolling scalar
avoided this problem but at the expense of having to account for  a
significant degree of fine tuning of parameters in the underlying
inflaton potential
\cite{Linde:1981mu,Albrecht:1982wi,Hawking:1981fz}. 

Over the past 20
years or so there have been many models of inflation, generally
implementing the idea of slow roll inflation
through the addition of extra fields. The best known of these, due to
Linde, is known as hybrid inflation \cite{Linde:1993cn}. However, it is
fair to say that in all these models there are parameters
which have to be fine tuned to the correct values. Even the most
recent ideas of inflation arising in brane worlds rely
on slow roll to provide the required amount of inflation
\cite{Dvali:1998pa,Steinhardt:2001st,Jones:2002cv,Kachru:2003sx} (For
nice reviews of inflation models in particle physics and string theory
see Refs.~\cite{Lyth:1998xn,Quevedo:2002xw}). 

One of the problems
facing models of inflation in the context of spontaneously broken
supergravity is that the moduli fields which would naively be
expected to be natural candidate inflaton fields generally fail to
satisfy the required slow roll condition $\eta \sim m^2/H^2 \ll 1$,
where $m$ is the mass of the field and $H$ the Hubble parameter during
inflation, since the fields generally have protected masses $m \sim
H$. This is the $\eta$-problem in inflation  and has proven to be a
real headache for those working in the field (see Ref.~\cite{Kachru:2003sx}
for a recent example, and 
Refs.~\cite{Kawasaki:2000yn,Hsu:2003cy,Koyama:2003yc,Firouzjahi:2003zy,%
Hsu:2004hi} for a recent proposal on how
a new shift symmetry in the K\"ahler potential could alleviate the
$\eta$-problem).

Dvali and Kachru \cite{Dvali:2003vv,Dvali:2003us}
recently revisited the idea of old inflation, asking
whether there was a way we could make use of the fact that it did not
rely on slow roll inflation, whilst avoiding the graceful exit
problems it usually has associated with it. 

They  proposed extending the
old inflation scenario, calling it {\it new old inflation} or {\it
locked inflation} (see also Ref.~\cite{Dimopoulos:2003ce}), and 
developed a model which did not
require a slow-roll potential, thereby alleviating a number of the
usual problems associated with inflation model building. It requires
two coupled fields, hence is in the spirit of hybrid inflation, but it
differs from the usual picture in that the universe arises from a
single tunnelling event as the inflaton leaves the false vacuum. The
subsequent dynamics arising from the oscillations of the inflaton
field keeps a second field trapped in a false minimum. For suitable
values of the parameters this then leads to a period of 50
e-foldings of inflation inside the bubble, allowing the bubble to grow
large enough to contain our present horizon volume. In this model,
reheating is accomplished when the inflaton driving the last stage of
inflation rolls down to the true vacuum, with the accompanying
adiabatic density perturbations arising from the moduli-dependent
Yukawa couplings of the inflaton to matter fields. In particular the
usual paradigm for calculating density perturbations from slow roll
inflation no longer applies and a formalism allowing for more general
scenarios has to be adopted 
\cite{Lyth:2001nq,Enqvist:2001zp,Dvali:2003em,Dvali:2003ar,%
Zaldarriaga:2003my,Kofman:2003nx}.

Following the work in Refs.~\cite{Dvali:2003vv,Dvali:2003us}, Easther 
et al.\ investigated some of the cosmological applications of locked
inflation \cite{Easther:2004qs}. They showed that an important
constraint on the allowed range of parameters on the model arises from
the fact that it is possible to have a secondary phase of {\it saddle
inflation} following  the initial period of locked inflation. This
subsequent period of inflation leads to a strongly scale dependent
spectrum and the possible overproduction of massive black holes in the
early universe. Avoiding this outcome places strong constraints on the
parameter space available to models of locked inflation.

In this paper, we extend the investigation of locked
inflation. As in Ref.~\cite{Easther:2004qs}, we concentrate on the
dynamics associated with the end of an extended period of new old
inflation, but we introduce new constraints emerging from
considerations of different physical properties associated with these
models. In general we find that it is very difficult to satisfy the
pure outcome associated with locked inflation, as there is invariably
some extra features which emerge to spoil the locked inflation. 

In
particular, adopting the model proposed in \cite{Dvali:2003vv}, we
consider the constraints which arise from density perturbations, loop
corrections, parametric resonance and defect formation at the end of
inflation. Our conclusions are quite strong. We show that  it is not
possible to satisfy all of the above constraints without having a
period of saddle inflation afterwards, which would wipe out all
observable signatures arising from the initial period of locked inflation
period. The one exception to this is that if we relax our insistence
that we satisfy the loop correction constraint. Then it is possible to
avoid a period of  saddle inflation,  but in that case we show that
the locked inflation must have ended through parametric resonance,
which may have important implications.

Following a discussion of the basics of the locked  inflation model in
Section~2, we begin to address the constraints imposed on the model in
Section~3. These include recapping the density perturbation constraint
arising from saddle inflation \cite{Easther:2004qs},
and demonstrating how this bound becomes much tighter once we account
for the fact that the period of inflation could end with the
production of topological defects. Introducing the quantum loop
corrections into the potential (expected to occur of course because
supersymmetry has to be broken today), we show that we are very far
from the slow roll regime. Turning our attention to the issue of
parametric resonance we show how the evolution equations for the
second waterfall field to which the inflaton is coupled, can be
written as a modified Mathieu equation. The evolution of this system
depends on the ratio of the mass scales of the inflaton ($m_\Phi$) and
waterfall field ($m_\phi$). The loop correction constrains this
parameter, $b \sim \frac{m_{\phi}^2}{m_{\Phi}^2} \leq 2$. We show
how the system behaves for large and small values of $b$,
corresponding to ignoring and satisfying the loop correction
constraint respectively. In particular, we show that in both cases,
inflation ends through parametric resonance. In Section~4, we
investigate the consequences of the parametric resonance for the
inhomogeneous modes, and show that in the case of small 
$b$, the non-linear effects associated with the resonance
rapidly cut off the inflation, meaning that no inflation
takes place. Finally we conclude in Section~5.

\section{Locked inflation}
The model discussed in Ref.~\cite{Dvali:2003vv}, is based on the same
potential as normal hybrid inflation~\cite{Linde:1993cn},
\begin{equation}
V(\Phi,\phi) =
\frac{1}{2}m_\Phi^2\Phi^2+\frac{1}{2}\lambda\Phi^2|\phi|^2+
\frac{\alpha}{4}\left(|\phi|^2-M_*^2\right)^2,
\end{equation}
where $\Phi$ and $\phi$ are scalar fields, possibly with multiple
components.  At the origin, the ``waterfall'' field $\phi$ has a
negative mass-squared $\mu^2=-m_\phi^2=-\alpha M_*^2$.  The potential
has symmetry-breaking minima at $|\phi|=M_*$ with $\Phi=0$.  

We assume that initially
$|\Phi|>\Phi_C=\sqrt{\alpha/\lambda}M_*$, so that the expectation value of
$\phi$ vanishes and the symmetry is restored.
The usual slow-roll parameter $\eta_\Phi$
is given by
\begin{equation}
\eta_\Phi=M_{\rm Pl}^2\frac{V''}{V}=\frac{m_\Phi^2}{3H^2},
\label{equ:etaPhi}
\end{equation}
where $M_{\rm Pl}=(8\pi G)^{-1/2}\approx 2.4\times 10^{18}$ GeV is the 
reduced Planck mass.
Slow roll inflation requires $\eta_\Phi\ll 1$, which 
poses a problem for attempts to build inflationary models from
string theory or supergravity, because in those cases $m_\Phi$ is  of
order $H$. This is known as the $\eta$-problem.  More generally, one
may ask how necessary slow roll is for inflation to occur. For these
reasons, it is interesting to study this model with $m_\Phi \gs H$.

In locked inflation, $m_\Phi > H$, but $|\Phi|$ is small enough so
that the energy density is dominated by the constant contribution
\begin{equation}
V_0=V(0,0)=\frac{\alpha}{4}M_*^4.
\end{equation}
This leads to an approximately exponential inflationary solution
with the expansion rate
\begin{equation}
H^2\approx
\frac{V_0}{3M_{\rm Pl}^2}=\frac{m_\phi^2M_*^2}{12M_{\rm Pl}^2}.
\end{equation}
At the same time,
$\Phi$ is oscillating with a decreasing amplitude
\begin{equation}
\Phi(t)\approx \Phi_0(t)\cos(m_\Phi^2-9H^2/4)^{1/2}t,
\end{equation}
where $\Phi_0(t)=\Phi_0\rme^{-3Ht/2}$.  

During every oscillation,
$\Phi$ passes through the instability region $|\Phi|<\Phi_C$, but if
its velocity is fast enough, it will leave it  before the instability
has had an effect. In Ref.~\cite{Dvali:2003vv}, the authors  assumed
that inflation goes on until the ``effective mass''
\begin{equation}
m_{\rm eff}^2\approx \lambda\Phi_0(t)^2-\alpha M_*^2,
\end{equation}
becomes negative. This happens when $\Phi_0(t)\approx \Phi_C$, and
implies that the number of e-foldings is
\begin{equation}
\label{equ:DKefold}
N=\frac{2}{3}\ln\frac{\Phi_0}{\Phi_C}\approx \frac{1}{3}
\ln\frac{\lambda\Phi_0^2}{m_\phi^2}.
\end{equation}

A different way to estimate when inflation ends was mentioned above
Eq.~(4) in Ref.~\cite{Dvali:2003vv}.  The time $\Phi$ spends in the
instability region  is approximately
\begin{equation}
\Delta t=\frac{2\Phi_C}{m_\Phi \Phi_0(t)}.
\end{equation}
This should be compared with the instability time  scale $1/m_\phi$,
which tells us that inflation ends when
\begin{equation}
\Phi_0(t)=\Phi_{C2}\approx  \frac{m_\phi}{m_\Phi}\Phi_C\approx
\frac{1}{\lambda^{1/2}} \frac{m_\phi^2}{m_\Phi}.
\label{equ:ownPhiend}
\end{equation}
As we will discuss later in
Section~\ref{sect:resonance}, this line of argument is the more appropriate
one in certain cases.

For the actual parameters used in Ref.~\cite{Dvali:2003vv}, the choice
between  $\Phi_C$ and $\Phi_{C2}$ makes no difference, because
$m_\Phi\approx m_\phi$.  The authors chose the ``natural'' values
$M_*\approx M_{\rm Pl}$ and $\lambda\approx 1$. They expressed the
other parameters $\alpha$ and $m_\Phi$ in terms of a mass scale $M$ by
writing $\alpha\approx M^4/M_{\rm Pl}^4$ and $m_\Phi^2\approx
M^4/M_{\rm Pl}^2$, from which the approximate equality between the two masses
follows.
The initial amplitude of the $\Phi$ field was chosen to be
$\Phi_0\approx M_{\rm Pl}$.

In this paper, we generally assume that $\lambda\sim 1$ 
and $\Phi_0\sim M_{\rm Pl}$. Because these parameter essentially
only enter the calculations logarithmically through Eq.~(\ref{equ:DKefold}),
their precise values are not very important.
For discussing the values of the remaining 
parameters, it turns out to be most useful to use the quantities
$\eta_\Phi$ defined in Eq.~(\ref{equ:etaPhi}), $\eta_\phi$ defined in 
Ref.~\cite{Easther:2004qs},
\begin{equation}
\eta_\phi= \frac{4M_{\rm Pl}^2}{M_*^2}=\frac{m_\phi^2}{3H^2},
\label{equ:etasmallphi}
\end{equation}
and the number of e-foldings $N$ calculated with Eq.~(\ref{equ:DKefold}),
which we can also write as
\begin{equation}
N\approx \frac{1}{3}\ln\frac{M_{\rm Pl}^2}{m_\phi^2}
\approx -\frac{1}{3}\ln\alpha+\frac{1}{3}\ln\frac{\eta_\phi}{4},
\end{equation}
and where the second term is generally negligible.
It can be seen that to have the $N\gs 50$ e-foldings required to
solve the flatness and horizon problems, one needs an incredibly
weak coupling $\alpha\sim 10^{-60}$ and a very low mass scale
$m_\Phi\sim m_\phi \sim 10^{-12}$\ GeV, but it is believed that
supersymmetry protects these from radiative
corrections~\cite{Dvali:2003vv}.

\section{Constraints on locked inflation}

\subsection{Density perturbations arising from the waterfall field}

Because the slow-roll conditions are not satisfied during inflation,
density perturbations cannot be created by the same mechanism as in
ordinary inflationary models. In Ref.~\cite{Dvali:2003vv}, the authors
proposed an alternative mechanism based on the modulus field that
controls the decay rate of $\phi$. However, it was later pointed out
in Ref.~\cite{Easther:2004qs}, that perturbations may be generated by
the usual slow-roll mechanism  if there is a period of ``saddle
inflation'' after the locked inflation.  If the period of saddle
inflation is too short, the spectrum of these  perturbations is
inconsistent with observations and may lead to production of massive 
black holes. Therefore only parameters for
which saddle inflation last for more than 50 e-foldings or does not
occur at all are possible. In the former case, saddle inflation would
wipe out all possible observable signatures of locked inflation, but
the latter case of no saddle inflation  is more interesting.

More specifically, the authors of Ref.~\cite{Easther:2004qs} 
calculated the amount of saddle inflation in terms of the 
parameter $\eta_\phi$ defined in Eq.~(\ref{equ:etasmallphi}).
This is the usual slow-roll parameter
in the $\phi$ direction (up to the sign), but even if it is greater
than unity, some amount of saddle inflation is possible.  
Defining a function
\begin{equation}
f(\eta_\phi)= \frac{3}{2}\left(\sqrt{1+\frac{4}{3}\eta_\phi}
-1\right)\approx \sqrt{3\eta_\phi},
\end{equation}
they showed that the number of e-foldings due to saddle inflation is
\begin{equation}
N_{\rm saddle}\approx\frac{1}{f(\eta_\phi)}
\ln\frac{\sqrt 2}{f(\eta)}\frac{M_{\rm Pl}}{H}
\approx \frac{1}{\sqrt{3\eta_\phi}}\ln\sqrt{2}\frac{M_{\rm Pl}}{m_\phi}.
\end{equation}
In terms of the number 
of e-foldings $N$ due to locked inflation, this is
\begin{equation}
N_{\rm saddle}\approx \sqrt{\frac{3}{4\eta_\phi}}N.
\end{equation}
In order to avoid saddle inflation, we need $N_{\rm saddle}\ls 1$,
and that gives the constraint
\begin{equation}
\eta_\phi\gs \frac{3}{4}N^2 \left( \approx 2000~\mbox{for $N=50$}\right).
\label{equ:etaconstraint}
\end{equation}

\subsection{Defect formation}

Because the end of inflation in this model involves a
symmetry-breaking phase transition, topological defects are produced
if the model allows them. This is the case if $\phi$ has fewer than
four real components.

For real $\phi$ the defects would be domain walls, which are known to
have disastrous cosmological consequences. They would soon dominate
the energy density of the universe, and cause it to collapse in a
short time.

If $\phi$ is complex, the defects are global strings. They have a
logarithmic long-range interaction, which is difficult to study
numerically, and  their effects are therefore not as well understood
as those of gauged strings. Nevertheless, it is believed that the
general consequences would be similar. In particular, for large enough
string tension they would produce observable temperature fluctuations
in the cosmic microwave background.  There is therefore an
observational upper limit for the tension $\mu$,  i.e., the  energy
per unit length, of a string of \cite{Jeong:2004ut}
\begin{equation}
G\mu\ls 10^{-6}.
\end{equation}
For global strings, $\mu$ is logarithmically diverging,
\begin{equation}
G\mu\approx \frac{M_*^2}{4M_{\rm Pl}^2}\log\frac{r}{r_0} =
\frac{1}{\eta_\phi}\log\frac{r}{r_0}.
\end{equation}
Thus, even if we ignore the logarithmic divergence, we have the
constraint $\eta_\phi\gs 10^6$, which is even stronger than that in
Eq.~(\ref{equ:etaconstraint}).

A very similar constraint applies if $\phi$ is an SO(3) triplet,
because in that case global monopoles are formed. Unlike gauged
't~Hooft-Polyakov  monopoles, they have a linearly increasing
interaction potential, and therefore they behave essentially as
endpoints of cosmic strings, and have similar cosmological
consequences.

Making the field $\phi$ charged under a gauge group would not relax
these constraints, because the constraint from the cosmic strings is
essentially the same in that case. In fact, formation of gauged
monopoles would only make things more difficult, because it would lead
to the well-known monopole problem.

Thus, we conclude that defect formation imposes a  constraint
$\eta_\phi\gs 10^6$, unless $\phi$ has at least four components, in
which case no defects exist. In Ref.~\cite{Dvali:2003vv}, it was
suggested that $\phi$ could be an SU(2) doublet field to avoid this
problem.

\subsection{Loop corrections from the waterfall field}

Because supersymmetry is broken, it cannot fully protect the potential
from radiative corrections. According to Ref.~\cite{Dvali:2003us}, the
one-loop correction from the $\phi$ field is
\begin{equation}
\Delta V_1\approx  \frac{m_\phi^2}{64\pi^2}\Phi^2\ln\frac{\Phi^2}{Q^2},
\label{equ:loopcorr}
\end{equation}
where $Q$ is some mass scale. A quadratic tree-level mass term can be
absorbed in the definition of $Q$.  The potential in
Eq.~(\ref{equ:loopcorr}) has a local minimum at some  point
$\Phi>\Phi_C$, unless
\begin{equation}
1+\ln\frac{\Phi_C^2}{Q^2}>0.
\label{equ:localmin}
\end{equation}
The field would then become trapped in this minimum and inflation would
never end. 
We must therefore satisfy Eq.~(\ref{equ:localmin}),
but this means that the
effective mass defined as the curvature of the effective potential 
satisfies
\begin{equation}
m_\Phi^{\rm eff}(\Phi)^2>\frac{m_\phi^2}{32\pi^2}\left(
2+\ln\frac{\Phi^2}{\Phi_C^2}\right).
\end{equation}
Using Eq.~(\ref{equ:DKefold}) one can express  the effective mass at
the time when inflation started in terms of the number of e-foldings,
\begin{equation}
m_\Phi^2(\Phi_0)\gs \frac{m_\phi^2}{32\pi^2}(2+3N)\approx
\frac{N}{100}m_\phi^2.
\label{equ:massratio}
\end{equation}
In terms of the slow-roll parameter $\eta_\Phi$ defined in
Eq.~(\ref{equ:etaPhi}), this can be written as
\begin{equation}
\eta_\Phi\gs \frac{N}{100}\eta_\phi
\left(\approx \frac{\eta_\phi}{2}~\mbox{for $N=50$}\right).
\label{equ:loopconstraint}
\end{equation}
Using Eq.~(\ref{equ:etaconstraint}), one finds that $\eta_\Phi\gs 1000$,
which means that we
are very far from the slow-roll regime.

\subsection{Parametric resonance}
\label{sect:resonance}
Let us now consider the equation of motion for the waterfall field
$\phi$,
\begin{equation}
\ddot\phi+3H\dot\phi+\left[\lambda\Phi(t)^2-m_\phi^2\right]\phi=0.
\end{equation}
Defining a new time variable $\tau=m_\Phi t$ and rescaling the field
as $\chi=\exp(3Ht/2)\phi$, we can write this as
\begin{equation}
\chi''+\left[2q(\tau)(1-\cos 2\tau)-b \right]\chi=0,
\label{equ:mathieu}
\end{equation}
where
\begin{equation}
b=\frac{m_\phi^2}{m_\Phi^2}+\frac{9}{4}h^2, \quad
q(\tau)=q_0\rme^{-3h\tau}, \quad
q_0=\frac{\lambda\Phi_0^2}{4m_\Phi^2},\quad
h=\frac{H}{m_\Phi}=\frac{1}{\sqrt{3\eta_\Phi}}.
\end{equation}
We can also write
\begin{equation}
b=\frac{\eta_\phi}{\eta_\Phi}\left(1+\frac{3}{4\eta_\phi}\right),
\end{equation}
which, together with the constraint (\ref{equ:etaconstraint}) implies
that $b\approx \eta_\phi/\eta_\Phi$.  The loop constraint
(\ref{equ:loopconstraint}) can then 
be written as $b\ls 100/N$,
and the saddle inflation constraint
(\ref{equ:etaconstraint}) as
\begin{equation}
h=\sqrt{\frac{b}{3\eta_\phi}}
\ls \frac{2\sqrt{b}}{3N}\left(
\approx 0.015\sqrt{b}\right).
\label{equ:hconstraint}
\end{equation}

The rescaled equation~(\ref{equ:mathieu}) is 
nothing but the Mathieu equation~\cite{ref:mathieu} with
time-dependent parameters $q=q(\tau)$ and $a=2q(\tau)-b$. Because $h$
is small, $q(\tau)$ is varying slowly, and we can assume that at any
given time, the evolution is well approximated by the ordinary Mathieu
equation with those parameters.

\subsubsection{Small $b$}

We will first assume that $b$ is of order one or smaller.
This is required by the loop correction 
constraint (\ref{equ:massratio}), which implies
that $b\ls 2$ for $N=50$ e-foldings.

\begin{figure}
\begin{center}
\begin{tabular}{lll}
(a)&(b)&(c)\cr \epsfig{file=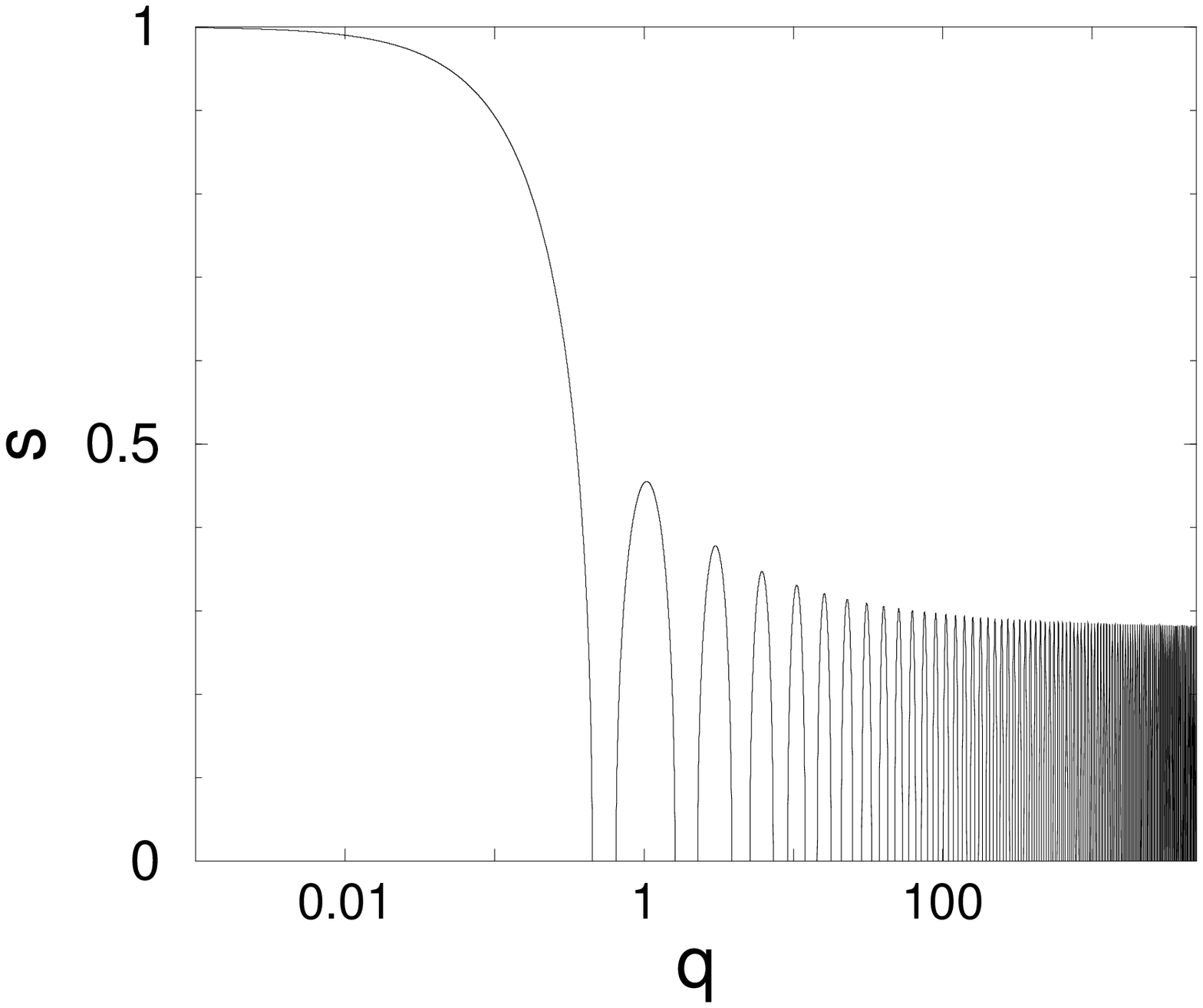,width=5cm}&
\epsfig{file=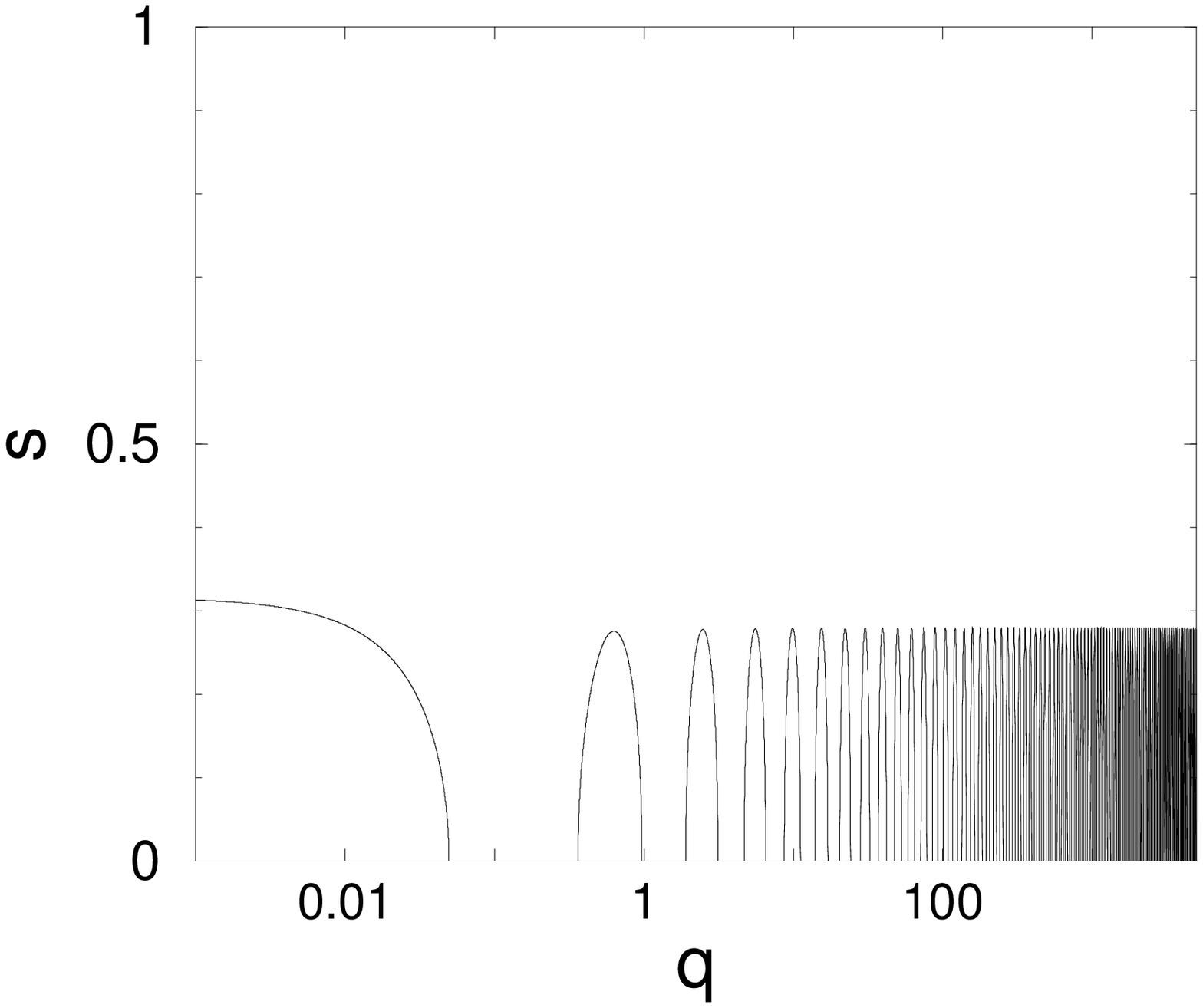,width=5cm}&
\epsfig{file=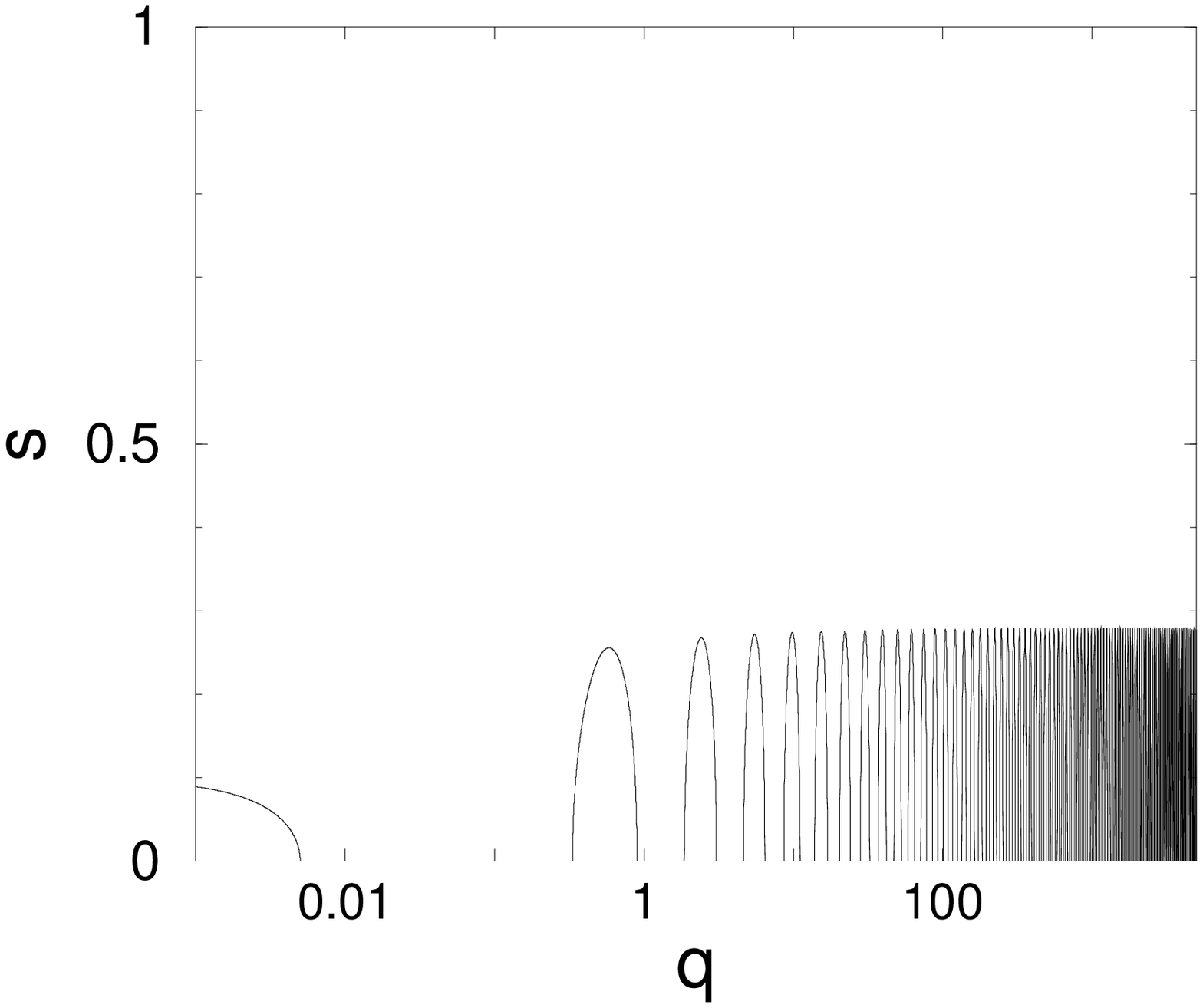,width=5cm}
\end{tabular}
\caption{
\label{fig:math1}
Floquet exponents for (a) $b=1$, (b) $b=0.1$ and (c) $b=0.01$.  }
\end{center}
\end{figure}

The Mathieu equation has stable and unstable regions in its parameter
space.  Floquet's theorem~\cite{ref:mathieu} shows
that the solutions are of the form
\begin{equation}
\chi(\tau)=\rme^{s\tau}f(\tau),
\end{equation}
where $f$ is periodic, $f(\tau+\pi)=f(\tau)$, and the Floquet exponent
$s$ can be complex. We calculated the exponent
numerically, by solving the equation in the interval $0\le\tau\le\pi$,
and have plotted the results for selected values of $b$ in
Fig.~\ref{fig:math1}.

\begin{figure}
\begin{center}
\epsfig{file=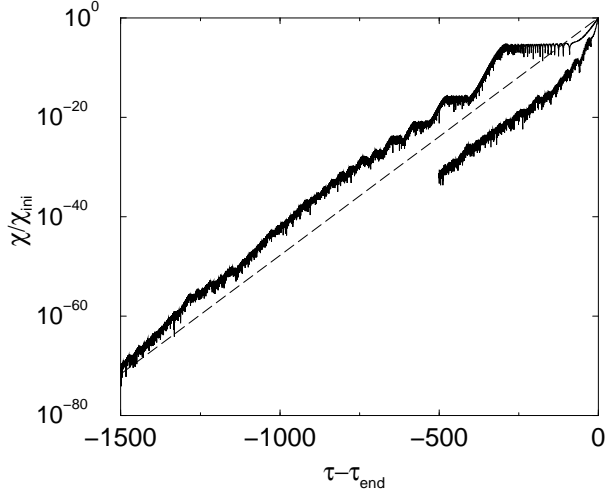,width=8cm}
\caption{
\label{fig:chi_smallb}
Numerical solutions of Eq.~(\ref{equ:mathieu}) for $(b=0.1,~h=0.003)$
(top curve) and $(b=1,~h=0.01)$ (bottom curve).  The dashed line is
the function $\exp[\bar{s}(\tau-\tau_{\rm end})]$  with
$\bar{s}=0.11$. In this plot, $\tau_{\rm end}$ is the time when the
evolution becomes tachyonic.  In other words,
$q(\tau)=(b/4)\exp[-3h(\tau-\tau_{\rm end})].$ }
\end{center}
\end{figure}

When $q\ls b/4$,  Eq.~(\ref{equ:mathieu}) is dominated by the constant
term $b$, and indeed, the plot shows that at those values, the system
undergoes a normal tachyonic instability. For higher $q$, there are
instability bands due to parametric resonance, and within each of
them, the maximum value of the exponent is around $0.3$. The mean
value of the exponent, averaged over a range of $q$, is approximately
$\bar{s}\approx 0.11$ for any $b\ls 1$.  We can therefore expect that
the solution of the full equation (\ref{equ:mathieu}) behaves as
$\chi(\tau)\sim\exp(\bar{s}\tau)$, implying that
\begin{equation}
\phi(t)\sim\rme^{(\bar{s}-3h/2)\tau}.
\label{equ:phigrowth}
\end{equation}
To check this assumption, we solved Eq.~(\ref{equ:mathieu})
numerically for two sets of parameters. As can be seen from
Fig.~\ref{fig:chi_smallb}, the results agree well with 
Eq.~(\ref{equ:phigrowth}).

According to Eq.~(\ref{equ:phigrowth}),  $\phi$ is growing
exponentially unless
\begin{equation}
h\gs 0.07
\left(\mbox{or equivalently $\eta_\Phi\ls 70$}\right).
\label{equ:resoconstraint}
\end{equation}
This constraint was mentioned on page 7 of Ref.~\cite{Dvali:2003vv}.
If we combine it with Eq.~(\ref{equ:hconstraint}), we find $b\gs N^2/100$,
which is only compatible with the original assumption of small $b$ 
if locked inflation lasts less than 10 e-foldings.
Parametric resonance is therefore inevitable
in a long period of locked inflation, if we want to satisfy both the 
saddle inflation and loop correction constraints.
The consequences of this will be discussed in section~\ref{ssec:smallb}.

\subsubsection{Large b}

One may also take the more phenomenological  view that it is not
necessary to satisfy the loop correction constraint, because the
underlying fundamental theory is unknown.  
Locked inflation then loses its main advantage over the usual slow-roll 
scenario, but it still remains an interesting alternative.
On the other hand, one can imagine a scenario with several shorter periods of
locked inflation~\cite{Dvali:2003vv}. 
In both cases $b$ can have larger values,
the behaviour 
of the Floquet exponent changes, and
Eq.~(\ref{equ:resoconstraint}) is not  valid.

\begin{figure}
\begin{center}
\epsfig{file=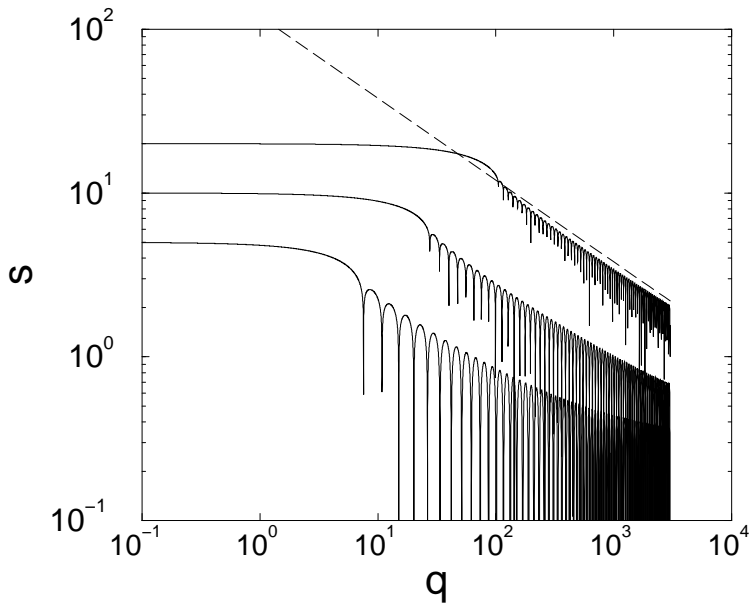,width=10cm}
\caption{
\label{fig:math2}
Floquet exponents for $b=400$, $b=100$ and $b=25$ (from top to
bottom).  The dashed line shows the curve $s=0.3b/\sqrt{q}$ for the
case $b=400$.  }
\end{center}
\end{figure}

In Fig.~\ref{fig:math2}, we show the Floquet exponent for larger
$b$. It can be seen from this curve that the exponent becomes large
well before the tachyonic range. Empirically, we find that the maximum
value of the exponent is well described by
\begin{equation}
s_{\rm max}\approx 0.3\frac{b}{\sqrt{q}}
\label{equ:smaxemp}
\end{equation}
for a wide range of $q\gs b$. This means that the resonance becomes
stronger well before the tachyonic range. In fact, the criterion
presented in Eq.~(\ref{equ:ownPhiend}) for the end of inflation can be
written as $ q_{\rm end}\approx b^2/4$, which is fully consistent
with Eq.~(\ref{equ:smaxemp}), indicating that inflation ends  when
$s_{\rm max}\approx 0.6$. However, this does not take into account the
expansion of the universe.

\begin{figure}
\begin{center}
\epsfig{file=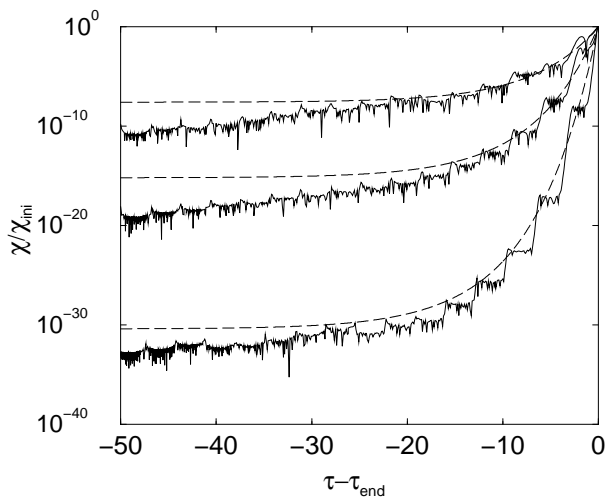,width=8cm}
\caption{
\label{fig:chi_largeb}
Numerical solutions of Eq.~(\ref{equ:mathieu}) for $b=25$, $b=100$ and
$b=400$ (from top to bottom) with $h=0.1$.  The dashed lines show the
growth predicted by the mean Floquet exponent $\bar{s}$ is
Eq.~(\ref{equ:barslargeb}).  In all three cases, the late-time
behaviour agrees with the prediction.  }
\end{center}
\end{figure}

Instead, we have to estimate whether the Hubble rate is high
enough to suppress this resonance.  Because the resonance bands are
broader for large $b$, the mean exponent $\bar{s}$ is greater than one
half of $s_{\rm max}$. By integrating the numerical solution for $s$
over $\tau$, we find that the mean value is approximately
\begin{equation}
\bar{s}(\tau)\approx 0.25\frac{b}{\sqrt{q(\tau)}}.
\label{equ:barslargeb}
\end{equation}
Again, we solved Eq.~(\ref{equ:mathieu}) numerically to check if it is
justified to use the mean exponent. Fig.~\ref{fig:chi_largeb} shows a
reasonably good agreement with the numerical data.  The resonance is
strong enough to overcome the Hubble  suppression when $\bar{s}\approx
3h/2$. If we assume that inflation ends then, the number of e-foldings
is
\begin{equation}
N'\approx \frac{1}{3}\ln\frac{36 h^2 q_0}{b^2} \approx \frac{1}{3}\ln
\frac{\lambda \Phi_0^2}{\alpha M_{\rm Pl}^2} \approx
N-\frac{1}{3}\ln\frac{\eta_\phi}{4},
\end{equation}
where $N$ is the number calculated from the tachyonic instability in
Eq.~(\ref{equ:DKefold}).
This logarithmic correction is typically small compared with $N$, so
the resonance does not reduce the amount of inflation significantly.
However, $N'<N$ for all realistic values of $\eta_\phi$, and therefore we can 
conclude that inflation always ends resonantly.

\section{Inhomogeneities under the influence of parametric resonance}

In Section~\ref{sect:resonance}, we showed that
that inflation always ends resonantly. For
small $b$, this happens because it is impossible to have a high enough
Hubble rate to suppress the resonant growth of $\phi$ without
violating Eq.~(\ref{equ:etaconstraint}). If we do not insist on the
loop correction constraint, $b$ can have larger values, but in that
case the Floquet exponent becomes so high at the late stages that the
resonance cannot be suppressed by any Hubble rate.

Let us now examine the consequences of this parametric resonance for
inhomogeneous modes. A mode with comoving momentum $k$ satisfies the
equation of motion
\begin{equation}
\chi_k''+\left[2q(\tau)(1-\cos 2\tau)-b(k,\tau) \right]\chi_k=0,
\label{equ:mathieuk}
\end{equation}
where $b(k,\tau)=b-\hat{k}^2\exp(-2h\tau)$ and $\hat{k}=k/m_\Phi$.

\subsection{Small $b$}
\label{ssec:smallb}

Let us first discuss the case with small $b\ls 1$. As long as
$q(\tau)\gs -b(k,\tau)$, the mode behaves in the same way as the
homogeneous mode, growing exponentially as
$\chi_k\sim\exp(\bar{s}\tau)$.

Because $q(\tau)$ is generally much larger than $b$, we can ignore the
constant in $b(k,\tau)$ and state that at any given time, modes with
\begin{equation}
\hat{k}^2\ls q_0\rme^{-h\tau}
\end{equation}
resonate. Because the right-hand side of this equation is decreasing
with  time, fewer and fewer comoving modes are resonating.

At time $\tau$, all modes with $k^2\ls k_{\rm max}^2(\tau)$, where
\begin{equation}
k_{\rm max}^2(\tau)\approx q_0m_\Phi^2\exp(-h\tau)
\approx\frac{\lambda \Phi_0^2}{4}\exp(-Ht) ,
\end{equation}
have been continuously amplified since the start of inflation.
The variance of those modes is the vacuum value multiplied by a
$k$-independent amplification factor,
\begin{equation}
\langle \chi_k^2\rangle(\tau) \approx \rme^{2\bar{s}\tau} \langle
\chi_k^2\rangle(0).
\label{equ:chiamp}
\end{equation}

Let us now consider the equation of motion for $\Phi$,
\begin{equation}
\ddot\Phi+3H\dot\Phi+\left(
m_\Phi^2+\lambda\langle\phi^2\rangle\right)\Phi=0.
\end{equation}
This equation becomes non-linear when the expectation value $\langle
\phi^2\rangle=\exp(-3Ht)\langle\chi^2\rangle$ exceeds $\langle
\phi^2\rangle_{\rm nl}\approx m_\Phi^2/\lambda$.  It is difficult to
solve the non-linear equation, so we will instead 
simply assume that inflation ends
when the non-linearity appears.  It would be interesting to carry out a more
detailed study of the non-linear dynamics to check that assumption,
but that is beyond the scope of this paper.

Assuming that all the modes with $k\ls k_{\rm max}$ are amplified by
the same amount, we can calculate the expectation value after $N$
e-foldings,
\begin{equation}
\langle\phi^2\rangle\approx \frac{k_{\rm max}^2}{8\pi^2}
\left[\rme^{(2\bar{s}/h-1)N}-1\right] \approx
\frac{\lambda\Phi_0^2}{32\pi^2}\rme^{-N}
\left[\rme^{(2\bar{s}/h-1)N}-1\right] ,
\label{equ:resovar}
\end{equation}
Because
the constant
prefactor in $\langle\phi^2\rangle$ is much greater than
$\langle\phi^2\rangle_{\rm nl}$, 
the evolution becomes non-linear during the first e-folding, and no
inflation will take place.
This rules out locked inflation with small $b$.

\subsection{Large $b$}

In the case of a larger $b\gg 1$, the Hubble rate can be higher and
suppress the resonance.  However, as Fig.~\ref{fig:chi_largeb}
shows, the amplification becomes extremely fast at the late stages.
We assume that Eq.~(\ref{equ:barslargeb}) is valid for time-dependent
$b$ in the form
\begin{equation}
\bar{s}(k,\tau)\approx 0.25\frac{b(k,\tau)}{\sqrt{q(\tau)}}.
\end{equation}
At the time when the resonance starts, i.e., $h\tau=N'$, inhomogeneous
modes start to resonate if $\hat{k}^2\exp(-2N')\ll b$. This includes
all superhorizon modes, for which $\hat{k}\exp(-N')\ls h$.

If the picture of amplification with the mean exponent $\bar{s}$ is
valid, all these modes grow very rapidly and inflation ends
instantaneously.  Therefore,  we do not expect any unwanted effects
such as density fluctuations on superhorizon scales or exceedingly strong
fluctuations on small scales. However,  a fully
non-linear study is needed to understand the details of the resonant
stage.

\section{Conclusions}

The prospect of avoiding the fine tuning associated with slow roll
inflation is very appealing. In their recent work, Dvali and Kachru
have  provided a mechanism  to do that by introducing a model
which effectively uses both the physics of old and new inflation
\cite{Dvali:2003vv,Dvali:2003us}. The motivation behind the proposal
is perhaps reinforced when we take on board the recent suggestions
concerning the string landscape
\cite{Bousso:2000xa,Feng:2000if,Maloney:2002rr,Kachru:2003aw,%
Susskind:2003kw,Ashok:2003gk,Acharya:2002kv,Acharya:2003gb}, 
which invokes the
possibility that there are a discrete set of closely spaced metastable
vacua in string theory. 

Following on from their work, Easther et 
al.~\cite{Easther:2004qs} 
pointed out a number of constraints which the model would have to
satisfy if its distinctive signatures were not to be completely
obliterated by a second extended period of saddle
inflation. In particular they showed that the
effective slow roll parameter associated with the second waterfall
field,  $\eta_\phi$ has to be huge, of order 1000, in order to prevent
this from occurring. 

In this paper we have also investigated the
dynamics of locked inflation in some detail, concentrating on some of
the non-linear features associated with the rich dynamics of the
coupled scalar fields present. 
We have
confirmed a number of the conclusions found in both
\cite{Dvali:2003vv} and \cite{Easther:2004qs} and have found some new
even stronger constraints. In particular we have have shown that the
whole parameter space available for locked inflation is ruled out if
we want to have a single, long period of inflation that would 
solve the flatness and horizon problems. This follows from the
following three observations:
\begin{enumerate}
\item Avoiding saddle inflation requires $\eta_\phi \gs 2000$
\item
Taking account of the loop correction requires $\eta_\Phi \gs
\eta_\phi/2$
\item Parametric resonance stops inflation during the
first e-folding unless $\eta_\Phi \ls 70$
\end{enumerate}

In addition we have shown the possible production of cosmic strings at
the end of the period of inflation places an even tighter constraint
on $\eta_\phi$, namely $\eta_\phi \gs 10^6$. However, this particular
constraint is avoided
if the waterfall field has more than three real components.

There are two possible ways of making the scenario viable. If one is willing
to have several shorter periods of inflation \cite{Dvali:2003vv}, 
the constraints become weaker. To illustrate this, we have calculated the 
same constraints for an arbitrary number of e-foldings.
Alternatively, a model with only one inflationary period is possible
if one does not insist on the loop correction constraint. In either case, our
results show that parametric resonance is inevitable. It is therefore
important 
to understand the consequences of this resonance to properly judge the
viability of locked inflation.

\section*{Acknowledgements}
EJC would like to thank Shamit Kachru for useful conversations when this work was being first started. AR was supported by Churchill College, Cambridge.


\section*{Bibliography}

\begin{thebibliography}{100}

\bibitem{Guth:1980zm}
A.~H.~Guth,
 ``The Inflationary Universe: A Possible Solution To The Horizon And Flatness
Problems,''
%
Phys.\ Rev.\ D {\bf 23}, 347 (1981).

\bibitem{Linde:1981mu}
A.~D.~Linde,
 ``A New Inflationary Universe Scenario: A Possible Solution Of The Horizon,
Flatness, Homogeneity, Isotropy And Primordial Monopole Problems,''
%
Phys.\ Lett.\ B {\bf 108}, 389 (1982).



\bibitem{Albrecht:1982wi}
A.~Albrecht and P.~J.~Steinhardt,
 ``Cosmology For Grand Unified Theories With Radiatively Induced Symmetry
Breaking,''
%
Phys.\ Rev.\ Lett.\  {\bf 48}, 1220 (1982).


\bibitem{Hawking:1981fz}
S.~W.~Hawking and I.~G.~Moss,
``Supercooled Phase Transitions In The Very Early Universe,''
Phys.\ Lett.\ B {\bf 110}, 35 (1982).

\bibitem{Linde:1993cn}
A.~D.~Linde,
``Hybrid inflation,''
Phys.\ Rev.\ D {\bf 49}, 748 (1994)
[arXiv:astro-ph/9307002].



\bibitem{Dvali:1998pa}
G.~R.~Dvali and S.~H.~H.~Tye,
``Brane inflation,''
Phys.\ Lett.\ B {\bf 450}, 72 (1999)
[arXiv:hep-ph/9812483].


\bibitem{Steinhardt:2001st}
P.~J.~Steinhardt and N.~Turok,
``Cosmic evolution in a cyclic universe,''
Phys.\ Rev.\ D {\bf 65}, 126003 (2002)
[arXiv:hep-th/0111098].

\bibitem{Jones:2002cv}
N.~Jones, H.~Stoica and S.~H.~H.~Tye,
``Brane interaction as the origin of inflation,''
JHEP {\bf 0207}, 051 (2002)
[arXiv:hep-th/0203163].


\bibitem{Kachru:2003sx}
S.~Kachru, R.~Kallosh, A.~Linde, J.~Maldacena, L.~McAllister and S.~P.~Trivedi,
``Towards inflation in string theory,''
JCAP {\bf 0310}, 013 (2003)
[arXiv:hep-th/0308055].



\bibitem{Lyth:1998xn}
D.~H.~Lyth and A.~Riotto,
``Particle physics models of inflation and the cosmological density
perturbation,''
Phys.\ Rept.\  {\bf 314}, 1 (1999)
[arXiv:hep-ph/9807278].

\bibitem{Quevedo:2002xw}
F.~Quevedo,
``Lectures on string / brane cosmology,''
Class.\ Quant.\ Grav.\  {\bf 19}, 5721 (2002)
[arXiv:hep-th/0210292].




\bibitem{Kawasaki:2000yn}
M.~Kawasaki, M.~Yamaguchi and T.~Yanagida,
``Natural chaotic inflation in supergravity,''
Phys.\ Rev.\ Lett.\  {\bf 85}, 3572 (2000)
[arXiv:hep-ph/0004243].



\bibitem{Hsu:2003cy}
J.~P.~Hsu, R.~Kallosh and S.~Prokushkin,
``On brane inflation with volume stabilization,''
JCAP {\bf 0312}, 009 (2003)
[arXiv:hep-th/0311077].



\bibitem{Koyama:2003yc}
F.~Koyama, Y.~Tachikawa and T.~Watari,
``Supergravity analysis of hybrid inflation model from D3-D7 system,''
Phys.\ Rev.\ D {\bf 69}, 106001 (2004)
[arXiv:hep-th/0311191].



\bibitem{Firouzjahi:2003zy}
H.~Firouzjahi and S.~H.~H.~Tye,
``Closer towards inflation in string theory,''
Phys.\ Lett.\ B {\bf 584}, 147 (2004)
[arXiv:hep-th/0312020].



\bibitem{Hsu:2004hi}
J.~P.~Hsu and R.~Kallosh,
 ``Volume stabilization and the origin of the inflaton shift symmetry in string
theory,''
%
JHEP {\bf 0404}, 042 (2004)
[arXiv:hep-th/0402047].


\bibitem{Dvali:2003vv}
G.~Dvali and S.~Kachru,
``New old inflation,''
[arXiv:hep-th/0309095].

\bibitem{Dvali:2003us}
G.~Dvali and S.~Kachru,
``Large scale power and running spectral index in new old inflation,''
[arXiv:hep-ph/0310244].


\bibitem{Dimopoulos:2003ce}
K.~Dimopoulos and M.~Axenides,
arXiv:hep-ph/0310194.

\bibitem{Lyth:2001nq}
D.~H.~Lyth and D.~Wands,
``Generating the curvature perturbation without an inflaton,''
Phys.\ Lett.\ B {\bf 524}, 5 (2002)
[arXiv:hep-ph/0110002].


\bibitem{Enqvist:2001zp}
K.~Enqvist and M.~S.~Sloth,
``Adiabatic CMB perturbations in pre big bang string cosmology,''
Nucl.\ Phys.\ B {\bf 626}, 395 (2002)
[arXiv:hep-ph/0109214].



\bibitem{Dvali:2003em}
G.~Dvali, A.~Gruzinov and M.~Zaldarriaga,
``A new mechanism for generating density perturbations from inflation,''
Phys.\ Rev.\ D {\bf 69}, 023505 (2004)
[arXiv:astro-ph/0303591].



\bibitem{Dvali:2003ar}
G.~Dvali, A.~Gruzinov and M.~Zaldarriaga,
 ``Cosmological Perturbations From Inhomogeneous Reheating, Freeze-Out, and 
Mass Domination,''
%
Phys.\ Rev.\ D {\bf 69}, 083505 (2004)
[arXiv:astro-ph/0305548].



\bibitem{Zaldarriaga:2003my}
M.~Zaldarriaga,
``Non-Gaussianities in models with a varying inflaton decay rate,''
Phys.\ Rev.\ D {\bf 69}, 043508 (2004)
[arXiv:astro-ph/0306006].



\bibitem{Kofman:2003nx}
L.~Kofman,
``Probing string theory with modulated cosmological fluctuations,''
arXiv:astro-ph/0303614.

\bibitem{Easther:2004qs}
R.~Easther, J.~Khoury and K.~Schalm,
``Tuning locked inflation: Supergravity versus phenomenology,''
JCAP {\bf 0406} (2004) 006
[arXiv:hep-th/0402218].

\bibitem{Jeong:2004ut}
E.~Jeong and G.~F.~Smoot,
``Search for cosmic strings in CMB anisotropies,''
arXiv:astro-ph/0406432.

\bibitem{ref:mathieu}
M.~Abramowitz and I.~A.~Stegun,
{\it Handbook of Mathematical Functions}, (Dover, New York, 1974).



\bibitem{Bousso:2000xa}
R.~Bousso and J.~Polchinski,
 ``Quantization of four-form fluxes and dynamical neutralization of the
cosmological constant,''
%
JHEP {\bf 0006}, 006 (2000)
[arXiv:hep-th/0004134].



\bibitem{Feng:2000if}
J.~L.~Feng, J.~March-Russell, S.~Sethi and F.~Wilczek,
``Saltatory relaxation of the cosmological constant,''
Nucl.\ Phys.\ B {\bf 602}, 307 (2001)
[arXiv:hep-th/0005276].



\bibitem{Maloney:2002rr}
A.~Maloney, E.~Silverstein and A.~Strominger,
``De Sitter space in noncritical string theory,''
arXiv:hep-th/0205316.



\bibitem{Kachru:2003aw}
S.~Kachru, R.~Kallosh, A.~Linde and S.~P.~Trivedi,
``De Sitter vacua in string theory,''
Phys.\ Rev.\ D {\bf 68}, 046005 (2003)
[arXiv:hep-th/0301240].



\bibitem{Susskind:2003kw}
L.~Susskind,
``The anthropic landscape of string theory,''
arXiv:hep-th/0302219.



\bibitem{Ashok:2003gk}
S.~Ashok and M.~R.~Douglas,
``Counting flux vacua,''
JHEP {\bf 0401}, 060 (2004)
[arXiv:hep-th/0307049].



\bibitem{Acharya:2002kv}
B.~S.~Acharya,
``A moduli fixing mechanism in M theory,''
arXiv:hep-th/0212294.



\bibitem{Acharya:2003gb}
B.~S.~Acharya,
``Compactification with flux and Yukawa hierarchies,''
arXiv:hep-th/0303234.


\end{thebibliography}
\end{document}